\documentclass[superscriptaddress,reprint,preprintnumbers,amsmath,amssymb,aps,prb,floatfix,
]{revtex4-2}

\usepackage{graphicx}
\usepackage{dcolumn}
\usepackage{bm}
\usepackage{hyperref}
\usepackage{amsmath,amssymb}
\hypersetup{colorlinks=true, linkcolor=blue, citecolor=blue, filecolor=blue,	urlcolor=blue}
\usepackage[normalem]{ulem}

\begin{document}
\title{
Origin of dimensional crossover in quasi-one-dimensional hollandite K$_{2}$Ru$_{8}$O$_{16}$
}

\author{Asif Ali}
\affiliation{Department of Physics, Indian Institute of Science Education and Research Bhopal, Bhopal 462066, India}

\author{Sakshi Bansal}
\affiliation{Department of Physics, Indian Institute of Science Education and Research Bhopal, Bhopal 462066, India}

\author{B. H. Reddy}
\affiliation{Department of Physics, Indian Institute of Science Education and Research Bhopal, Bhopal 462066, India}%

\author{Ravi Shankar Singh}
\email{rssingh@iiserb.ac.in}
\affiliation{Department of Physics, Indian Institute of Science Education and Research Bhopal, Bhopal 462066, India}

\date{\today}

\begin{abstract}

Intriguing phenomenon of dimensional crossover  is comprehensively studied by experimental and theoretical investigation of electronic structure in quasi-one-dimensional hollandite K$_{2}$Ru$_{8}$O$_{16}$. Valence band photoemission spectra in conjunction with density functional theory within local density approximation combined with dynamical mean field theory (LDA+DMFT) reveal moderately correlated electronic structure. Anomalous temperature dependence of high-resolution spectra in the vicinity of Fermi level suggests Tomonaga-Luttinger liquid state down to 150 K, below which it undergoes a dimensional crossover from one-dimensional to three-dimensional electronic behaviour. Monotonously decreasing spectral intensity at the Fermi level along with Fermi cut-off at low temperature suggests non-Fermi liquid like behaviour. 
Many body effects captured within LDA+DMFT reveal increased warping of the Fermi surface with lowering temperature. A simple analysis suggests that the warping dominates the thermal energy induced momentum broadening at low temperature, leading to the 3D electronic behaviour. 
Our results offer valuable insight in  understanding the interplay of dimensionality, electron correlation and thermal energy governing various exotic phenomena in quasi-one-dimensional systems.

\end{abstract}

\maketitle
\section{Introduction}
Landau's Fermi liquid theory, a cornerstone theory to describe interacting electrons in ordinary metals, breaks down in one-dimension (1D) since the individual particle excitations become unstable due to strong interactions and the theory is replaced by the Tomonaga-Luttinger liquid (TLL) theory
\cite{Tomonage1950, Luttinger1963}. The TLL state has been experimentally realized in quantum wires \cite{ 1dnanowires_ESlot_PRL2004, Blumenstein2011}, carbon nanotubes \cite{Bockrath_CNT1999}, ultracold atoms \cite{RevModPhys.80.885} and 1D crystals \cite{PhysRevLett.88.096402, Nicholson_PRL2017}. {Going to 2D/3D structures by coupling 1D chains can lead to non-trivial physics \cite{RevModPhys.91.041001}.}
There are naturally occurring 3D materials containing weakly coupled linear chains that exhibit highly anisotropic properties due quasi-1D electronic structure \cite{PhysRevLett.95.186402}. However, the weak transverse (interchain) interaction is of great importance as it plays a decisive role in their physical properties \cite{PhysRevLett.98.266405, PhysRevLett.72.3218, Giamarchi_1dBook_2003}. An intriguing phenomenon of dimensional crossover is observed in many quasi-1D materials, essentially a system of coupled chains to decoupled chains, above a characteristic temperature $T^*$ \cite{Giamarchi_1dBook_2003, silke_PRL2004}.
This phenomenon is understood in terms of competition between transverse interactions leading to warping of the Fermi surface (FS) and the thermal energy, where the former becomes relevant below $T^*$, and the 1D behaviour is destroyed \cite{Giamarchi_1dBook_2003, silke_PRL2004}. Understanding of such dimensional crossover is crucial to explore the exotic properties realized in quasi-1D materials.    

Owing to the interplay of confinement and electron correlation among $d$ electrons, quasi-1D transition metal oxides (TMOs) are at the forefront of research on low-dimensional systems \cite{yamaura_NaV2O4_PRL2007, PhysRevLett.77.4054,PhysRevB.66.075107}. 
Hollandites are a class of TMOs having chemical formula $A_x$$M_8$O$_{16}$ ($A$ = alkali/alkaline-earth metals, $M$ = transition metal) \cite{Hollandite_review2019}. Their structure consists of edge-shared $M$O$_6$ octahedra forming $M$O$_2$ zig-zag chain along $c$-axis that connects with neighbouring chains to form a tunnel structure which accommodates $A$ ions (see Fig. S1 of Supplemental Material (SM) \cite{SM}).
This unique structure leads to novel phenomena, for example, metal-insulator transition in $A_x$V$_8$O$_{16}$ \cite{Isobe_MIT_JPSJ2006, Hesagawa_PRL2009, Maignan_PbVO_MIT_PRB2010}; Peierls Mott-Insulator transition in ferromagnetic K$_2$Cr$_8$O$_{16}$ \cite{Bhobe_PRX2015}; charge and orbital ordering in Ba$_x$Ti$_8$O$_{16}$ and K$_2$V$_8$O$_{16}$ \cite{komarek_ChargeOrder_PRL2011,BTO_ordering_PRB2015}; correlated metallic state in K$_x$Ir$_8$O$_{16}$ \cite{IrHollandite_InorgCHem2014}; and dimensional crossover along with non-Fermi liquid behaviour in Ba$_{1.2}$Rh$_8$O$_{16}$, Ba$_{1.33}$Ru$_8$O$_{16}$ and  K$_2$Ru$_8$O$_{16}$ \cite{RhHollandite_APautrat_PRB2010, Cava_QPT_BRO_PRL2003, kobayashi_KROPRB2009, *Pautrat_2012} \textit{etc}. 
Among these, 4$d$ ruthenium hollandite K$_{2}$Ru$_{8}$O$_{16}$ exhibits anisotropic quasi-1D behaviour and has been proposed to host TLL state \cite{kobayashi_KROPRB2009, TToriyama_KRO_PRB2011}. Temperature dependent resistivity exhibits metallic behaviour along the tunnel direction (\textit{c}-axis), while a broad insulator-metal transition is observed along the transverse direction (\textit{ab}-plane) associated with the dimensional crossover at $\sim$ 150 K \cite{kobayashi_KROPRB2009,*Pautrat_2012}. Non-Fermi liquid behaviour has been observed in low temperature resistivity along with signature of moderate electron correlation as indicated by Sommerfeld coefficient $\gamma$ = 24~mJ~mol$^{-1}$~K$^{-2}$ from specific heat measurements which is 1.7 times larger than that estimated from density functional theory (DFT) ($\gamma_{DFT}$ = 14 mJ~mol$^{-1}$~K$^{-2}$) \cite{FOO_KRO2004}. 
{This suggests that calculations beyond DFT might be required to accurately describe the electronic structure}.
However, this mass enhancement observed in K$_2$Ru$_8$O$_{16}$ is smaller than various other ruthenates
\cite{GCao_CSRO_PRB1997, Oguchi_SRO214_PRB1995, PhysRevB.62.R6089}. No structural distortion/transition or long range magnetic order has been observed down to very low temperature \cite{FOO_KRO2004, kobayashi_KROPRB2009, *Pautrat_2012} 
making K$_2$Ru$_8$O$_{16}$ a suitable candidate to explore the physics of dimensional crossover in quasi-1D systems. 

Here, we elucidate the role of electron correlation and thermal energy on the dimensional crossover in the quasi-1D hollandite, K$_{2}$Ru$_{8}$O$_{16}$. Signature of lower Hubbard band along with renormalized Ru 4$d$ band in the valence band photoemission spectra, indicating moderate electron correlation, are well captured by DFT within local density approximation in combination with dynamical mean field theory (LDA+DMFT). High resolution spectra in the vicinity of Fermi energy ($E_F$) exhibit an anomalous temperature dependence, where the spectral intensity follows TLL lineshape suggesting 1D behaviour down to 150 K and deviate at low temperatures due to the dimensional crossover to the higher dimensional behaviour, commensurate with the transport measurements. The dimensional crossover is further corroborated by temperature dependent LDA+DMFT calculations. 

\section{METHODOLOGY}
Polycrystalline sample of K$_{2}$Ru$_{8}$O$_{16}$ was synthesized by solid state reaction method. Stochiometric mixture of high purity RuO$_{2}$ (99.99 \%) and K$_{2}$CO$_{3}$ (99.99\%) was thoroughly grounded and heated in palletized form at 1000 $^{\circ}$C for 15 hours under nitrogen gas flow. An additional 15$\%$ K$_{2}$CO$_{3}$ was used to compensate for potassium evaporation. Room temperature powder $x$-ray diffraction  pattern was collected using Cu $K_{\alpha}$ radiation to analyze the phase purity and crystal structure of the prepared sample (see Note S2 of SM \cite{SM}).
Temperature dependent photoemission spectroscopic measurements were performed using Scienta R4000 electron energy analyzer and monochromatic $x$-ray and ultra-violet radiations. Multiple samples were fractured \textit{in-situ} under ultra high-vacuum (base pressure $\sim$ 5 $\times$ 10$^{-11}$ mbar) to obtain clean sample surface. The Fermi level ($E_{F}$) and total energy resolution were determined by measuring the Fermi edge of a clean polycrystalline Ag. The experimental energy resolutions were set to 300 meV, 5 meV and 10 meV for Al $K_{\alpha}$ ($h\nu$ = 1486.6 eV), He~\textsc{i} ($h\nu$ = 21.2 eV) and He~\textsc{ii} ($h\nu$ = 40.8 eV) radiations (photon energy), respectively.  

Electronic structure calculations were performed within density function theory (DFT) framework using full potential linearized augmented plane wave method as implemented in \textsc{wien2k} with local density approximation (LDA) for the exchange-correlation functional \cite{WIEN2k, *Wien2k2019}. Fully relaxed lattice parameters and atomic positions were adopted for all the calculations (force convergence $<$ 10$^{-3}$ eV \AA $ ^{-1}$). The muffin tin radii for K, Ru and O were set to 2.35, 1.93, and 1.66 Bohr radius, respectively. The size of the basis set was determined by $R_{mt}K_{max}$ = 7.0, where $R_{mt}$ and $K_{max}$ stand for smallest muffin-tin radii and largest plane wave vector, respectively.  The Brillouin zone integration was performed using a 14 $\times$ 14 $\times$ 14 \textit{k}-mesh with energy and charge convergence criteria set to 10$^{-5}$ eV and 10$^{-4}$ electronic charge per formula unit, respectively. The LDA+$U$ method \cite{AnisimovLDAU} was adopted to account for on-site electron correlation for the Ru 4$d$ orbital.

Fully charge self-consistent LDA+DMFT calculations were performed using eDMFT code \cite{HauleDMFTPRB2010} combined with \textsc{wien2k} \cite{WIEN2k, *Wien2k2019}. The continuous time quantum Monte Carlo (CTQMC) method was employed for impurity solver \cite{CTQMC_PhysRevLett.97.076405, Haule_CTQMC_PRB2007}. Projectors within $\pm$ 10 eV energy window around Fermi level were used to construct local orbitals basis where all five Ru 4$d$ orbitals were considered as correlated. The \textit{``exact''} method was used for double-counting corrections \cite{HauleExactDCPRL2015}. Rotationally invariant form of the Coulomb interaction was used. In order to calculate spectral functions, analytical continuation of the self-energy from imaginary to real frequency axis was performed using maximum entropy method \cite{JARRELL1996133}. The Fermi surface was calculated on a denser 31 $\times$ 31 $\times$ 31 $k$-mesh.
The momentum dependent quasi-particle spectral function $A(k,\omega)$ is described by 
    \begin{equation}\label{eq:1}
	A(\textit{\textbf{k}},\omega) = -\frac{1}{\pi} \frac{Im~\Sigma(\omega)}{[\omega + \mu - \epsilon_\textit{\textbf{k}} - Re~\Sigma(\omega) ]^2 + [Im~\Sigma(\omega)]^2}
\end{equation} 
where, $\omega$ is the real frequency, $\mu$ is chemical potential, $Re~\Sigma(\omega)$ ($Im~\Sigma(\omega)$) represent the real (imaginary) part of the self-energy on the real-frequency axis and $\epsilon_{\textit{\textbf{k}}}$ represents the eigenvalue of the non-interacting electron (LDA) with crystal momentum $k$.  
Mass enhancement factor ($m^*/m_{b}$) was obtained as weighted sum of contributions arising from all the orbitals (estimated by 1 $-$ $\partial \text{Im}[\Sigma(i\omega)] / \partial (\omega)|_{\omega \rightarrow 0}$) and weighted by their local Green's function ($\propto$ partial DOS) at $E_F$ \cite{GKotliar_RevModPhys.78.865}, where Im$[\Sigma(i\omega)]$ is imaginary part of the self-energy on imaginary frequency axis.  Mass enhancement factor was also obtained from the slopes of real part of the  self-energy on real frequency axis Re$[\Sigma(\omega)]$, and were consistent.

\section{Results and Discussion}
\begin{figure}[t]
	\includegraphics[width=.48\textwidth, trim= 0 0 0 0,clip]{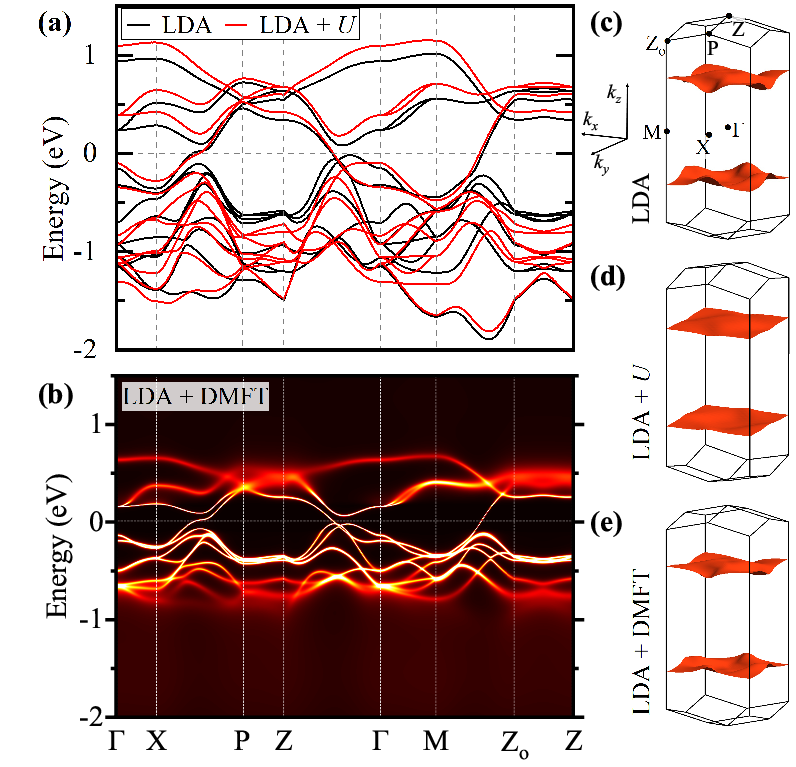}
	\caption{ (a) Band dispersion along high-symmetry direction obtained using LDA (black lines) and LDA+$U$(= 3.3 eV) (red lines) calculations for K$_{2}$Ru$_{8}$O$_{16}$. Fermi surfaces obtained using (c) LDA and (d) LDA+$U$. (b) Momentum resolved spectral function {$A(k,\omega$)} and (e) Fermi surface obtained from LDA+DMFT calculations at 100 K. High symmetry points have been labeled in (c).}\label{fig:Fig1}
\end{figure}
We begin with a basic description of the electronic structure of K$_{2}$Ru$_{8}$O$_{16}$, where Fig.~\ref{fig:Fig1}(a) shows band structure around $E_F$ calculated within LDA (black lines). The bands between -2 eV and 1 eV energy region have predominantly Ru 4$d$ contribution, whereas the well separated O 2$p$ band mainly appears below -2.6 eV energy \cite{TToriyama_KRO_PRB2011}. Notably, only one band crosses $E_F$ in the $\Gamma$-Z, M-Z$_{0}$ and X-P directions and two parallel sheets like FS is obtained, as shown in Fig.~\ref{fig:Fig1}(c).
The finite warping of the FS is caused by finite transverse interactions orthogonal to the tunnel direction (carrier hopping between the zig-zag chains). The warping is proportional to transverse hopping: zero warping would mean effectively decoupled chains, thus a perfect 1D conductor, while larger warping would mean increased interchain interaction leading to higher dimensional behaviour. Therefore, the effective dimensionality of a quasi-1D system can be speculated from the warping of the FS. 

A moderate strength of electron correlation is indicated by mass enhancement factor of 1.7 estimated from specific heat measurement \cite{FOO_KRO2004}.    
We obtained an average effective screened Coulomb interaction $U_{cRPA}$ $\approx$ 3.3 eV using the constrained random phase approximation (cRPA) calculation (details in Note S1 of SM \cite{SM}). The inclusion of electron correlation within LDA+$U$ framework \cite{AnisimovLDAU} ($U$ = $U_{cRPA}$) pushes the states away from $E_{F}$ and moderately increases the width of Ru 4$d$ band (red lines in Fig. \ref{fig:Fig1}(a)). The band dispersion close to $E_F$ is slightly modified, leading to significantly reduced warping of the FS, as shown in Fig. \ref{fig:Fig1}(d), suggesting highly 1D character. This, however, is in contrast with the finite transport observed along transverse direction (\textit{ab}-plane) \cite{kobayashi_KROPRB2009}. 
The LDA+$U$ method works well in strongly correlated systems, whereas the case of weak/moderate correlation is more challenging, for which the LDA+DMFT method has emerged as very successful technique \cite{Anisimov_1997, GKotliar_RevModPhys.78.865, RuthenatesDMFT_PRL2016, IridatesDMFT_PRL2013, SRO_MinjaeKIM,  SRO_SRARPES_PRL2021, Ali_SMO_JPCM2023, APaulPRB2019}.  
In Fig.~\ref{fig:Fig1}(b), we further show the momentum resolved spectral function obtained from LDA+DMFT at $\beta$ = 116 eV$^{-1}$ ($T \approx$ 100 K) for $U$ = 5.0 eV and $J$ = 0.5 eV. Similar values have also been used for other 4$d$ systems \cite{APaulPRB2019, Ali_SMO_JPCM2023}. 
The overall spectral function exhibits renormalized Ru 4$d$ $t_{2g}$ bands, which appear within -1 eV to 0.75 eV along with highly diffused incoherent bands appearing at higher energies away from $E_{F}$. The obtained mass enhancement factor of about 1.5 is in good agreement with the estimated value from specific heat \cite{kobayashi_KROPRB2009}. 
Surprisingly, the LDA+DMFT FS exhibits substantial warping similar to that of the LDA FS, as shown in Fig.~\ref{fig:Fig1}(e). 

\begin{figure}[t]
	\includegraphics[width=.45\textwidth, trim= 0 0 0 0,clip]{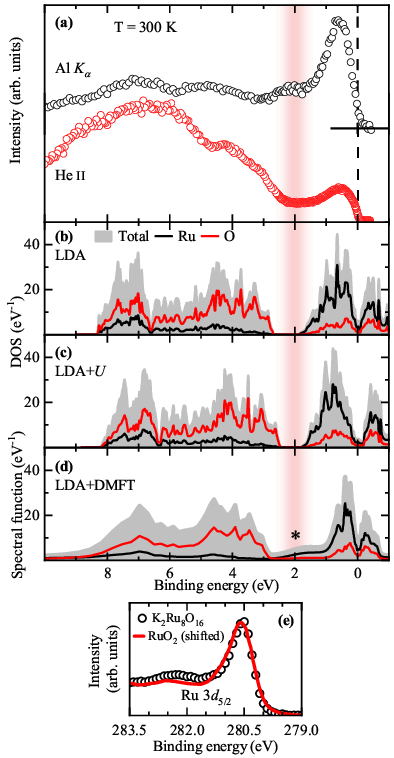}
	\caption{(a) Valence band photoemission spectra of K$_{2}$Ru$_{8}$O$_{16}$ collected at 300 K using Al $K_\alpha$ (black) and He~\textsc{ii} (red) radiations. Total and atom resolved DOSs calculated within (b) LDA and (c) LDA+$U$. (d) Total and atom resolved spectral functions obtained within LDA+DMFT at 100 K, where asterisk marks the incoherent feature. Vertically shaded region around 2 eV binding energy highlights the incoherent feature. (e) Ru 3$d_{5/2}$ core level spectra for K$_{2}$Ru$_{8}$O$_{16}$ (markers) and RuO$_2$ (line). The RuO$_2$ spectra (from Ref.\cite{Ru3d_DJMorgan}) has been shifted towards lower binding energy by 0.25 eV.}\label{fig:Fig2}
\end{figure}
To get valuable insight about the role of electron correlation and to validate the suitability of above mentioned theoretical frameworks, we present the valence band photoemission spectra of K$_{2}$Ru$_{8}$O$_{16}$ collected at 300 K using Al $K_{\alpha}$ and He~\textsc{ii} radiations in Fig. \ref{fig:Fig2}(a). 
Both the spectra exhibit a prominent feature at around 0.5 eV binding energy and multiple broad features beyond $\sim$ 1.5 eV binding energy. The intensity of features beyond $\sim$ 3 eV binding energy is relatively larger in He~\textsc{ii} spectra, while the features below $\sim$ 3 eV binding energy are significantly enhanced in the Al $K_{\alpha}$ spectra. This dependence of relative spectral intensity on the incident photon energy can be primarily associated with photo-ionization cross-section as the cross-section ratio for O 2$p$ states to Ru 4$d$ states is larger at lower photon energies \cite{YEH19851}. Thus, the features above $\sim$ 3 eV binding energy can be primarily attributed to O 2$p$ states, while the lower binding energy features predominately have Ru 4$d$ character, also consistent with the atom-resolved density of states (DOSs) calculated within LDA, as shown in  Fig. \ref{fig:Fig2}(b).
A fairly good agreement between LDA and experimental results is evident; for example, a dip like structure in LDA DOS commensurate with decreasing photoemission intensity in the vicinity of $E_F$ along with various feature positions and total width of O 2$p$ band.
However, the occupied Ru 4$d$ band within LDA appears well below 2 eV binding energy, while the experimental spectra exhibit an additional feature at 2 eV binding energy, highlighted by the vertically shaded region. The additional feature (absent in LDA) is the signature of correlation induced lower Hubbard band (incoherent feature), which has also been observed in various other ruthenates \cite{CaSrRuO3_RSSINGH2005, *SROTakizawa_PRB2005, Horio2023, PhysRevLett.96.107601}. As shown in Fig. \ref{fig:Fig2}(c), the inclusion of electron correlation $U$ = 3.3 eV within LDA+$U$ also fails to reproduce this additional feature. We note that flat FS obtained within LDA+$U$ (Fig. \ref{fig:Fig1}(d)) is in disagreement with finite transverse transport. 
On the other hand, diffusive bands obtained within LDA+DMFT (Fig. \ref{fig:Fig1}(b)) lead to a moderately intense incoherent feature appearing at $\sim$ 2 eV binding energy (marked by asterisk), as shown in Fig. \ref{fig:Fig2}(d). Thus, the LDA+DMFT successfully describes the photoemission spectra capturing the O 2$p$ band, the renormalized Ru 4$d$ band along with the correlation induced incoherent feature (see appendix \ref{sec:appendixA} for varying $U$ and $J$ values).
These results establish the importance of dynamical correlation accounted within LDA+DMFT framework to capture the many-body effects in the electronic structure of correlated  K$_{2}$Ru$_{8}$O$_{16}$.

The signature of electron correlation in ruthenates has also been identified as appearance of the satellite features in the Ru 3$d$ core level spectra \cite{HDKim_Ru_PRL2004, SBansal_LRO_JPCM2022}. The Ru 3$d_{5/2}$ core level spectra exhibits a weak unscreened feature appearing at $ \sim $ 282.2 eV binding energy along with the screened feature at $ \sim $ 280.5 eV binding energy for K$_2$Ru$_8$O$_{16}$, as shown in Fig. \ref{fig:Fig2}(d). Comparison of Ru 3$d$ core level spectra reveals a shift of about 0.25 eV towards lower binding energy in K$_2$Ru$_8$O$_{16}$ with respect to RuO$_2$ \cite{Ru3d_DJMorgan}, suggesting a rigid band shift due to nominal Ru$^{3.75+}$ oxidation state in the former compared to Ru$^{4+}$ oxidation state in the latter. Moderately larger intensity of the unscreened feature in K$_2$Ru$_8$O$_{16}$ as compared to RuO$_2$ also indicates moderate strength of electron correlation in this system.

\begin{figure}
	\includegraphics[width=.48\textwidth]{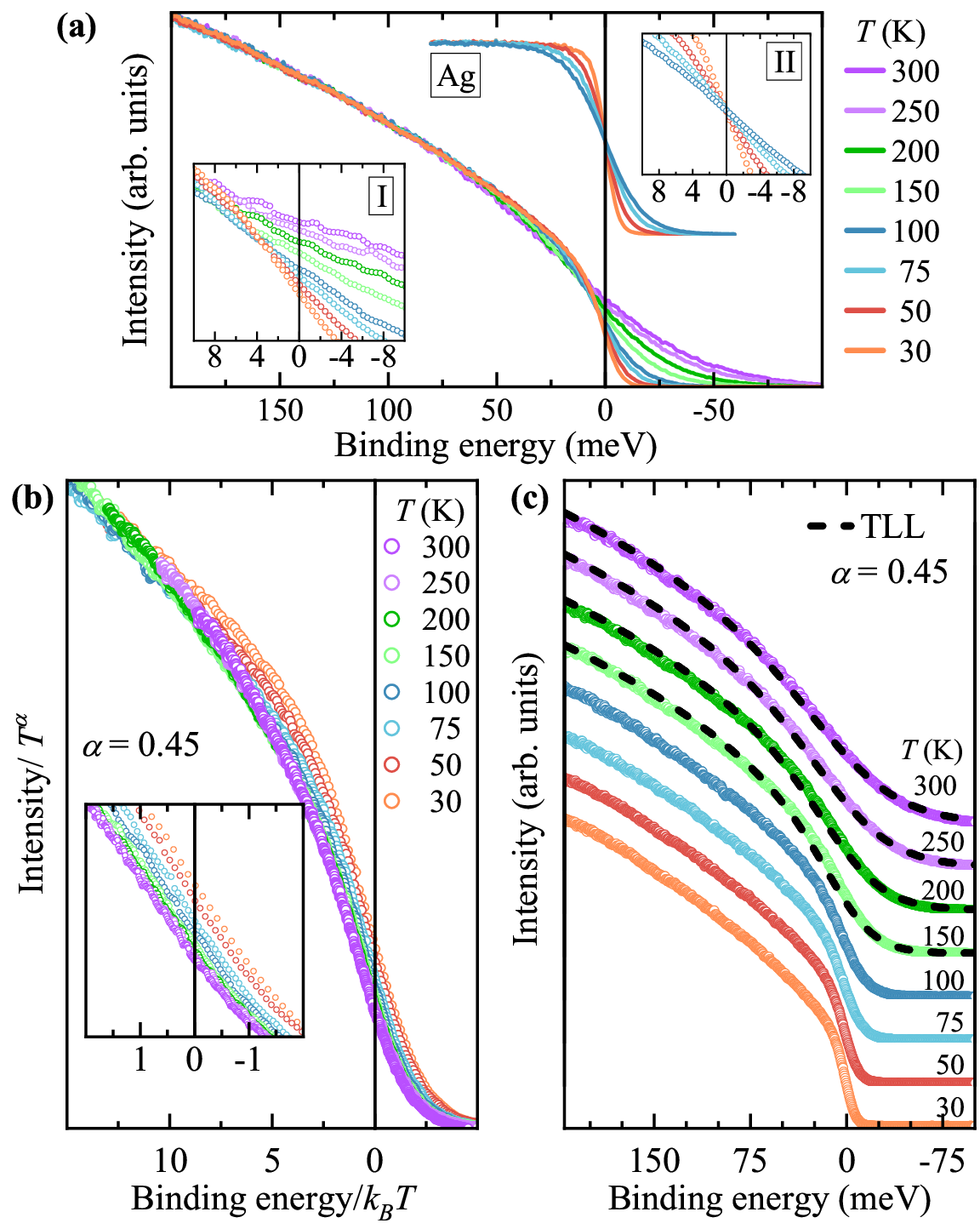}
	\caption{(a) Temperature dependent high-resolution photoemission spectra of K$_{2}$Ru$_{8}$O$_{16}$ collected using He~\textsc{i} radiation. The inset I shows the monotonously decreasing spectral intensity at $E_{F}$. The temperature dependent spectra for polycrystalline Ag is also presented (vertically shifted for clarity) and inset II shows the Ag spectral intensity close to $E_F$. (b) Scaled photoemission spectral intensity for $\alpha$ = 0.45. Inset shows the deviation of spectra at $T \leq$ 100 K in expanded scale. (c) Comparison of photoemission spectra (stacked vertically) with TLL lineshape at different temperatures. A clear Fermi cut-off is observed at low temperature.}\label{fig:Fig3}
\end{figure}   
    
Although the effects of electron correlation are manifested in the large energy scale as described above, various thermodynamic properties are essentially determined by the energy states near $E_F$ ($|E-E_F|$ $\approx$ $k_BT$). Thus, to further understand the anomalous transport behaviour suggesting dimensional crossover and non-Fermi liquid behaviour, we show the high resolution temperature dependent valence band spectra in the vicinity of $E_{F}$ using He~\textsc{i} radiation in Fig. \ref{fig:Fig3}(a). Normalization of all the spectra at around 200 meV binding energy shows similar lineshape down to 75 meV binding energy at all temperatures. The spectra in the energy range close to $E_F$ reveal interesting evolution along with appearance of Fermi cut-off at low temperature. The monotonously decreasing spectral intensity at $E_F$ is evident in inset I. For comparison, we show the near $E_F$ spectra of polycrystalline Ag (vertically shifted for clarity) exhibiting Fermi-Dirac like temperature evolution with no change of the spectral intensity at $E_F$ (see inset II). The monotonously decreasing states at the Fermi energy with lowering temperature have been observed in a variety of systems exhibiting various phase transitions and/or anomalous transport behaviours \textit{e.g.} pseudo-gap phase in superconductors, charge density wave (CDW) in low dimensional systems, electronic/structural/magnetic phase transitions leading to metal-insulator transition, disordered systems and 1D/quasi-1D metals following TLL behaviour \cite{Vishik_2018, MgB2_Patil2017, PhysRevLett.81.2974, RevModPhys.70.1039, RevModPhys.57.287,  Wang_LiMoO_PRL2006,PhysRevLett.82.2540,Kang2021}

Various bulk measurements such as specific heat, magnetization and electrical transport have revealed that K$_{2}$Ru$_{8}$O$_{16}$ remains Pauli paramagnetic metal down to very low temperatures with no signature of superconductivity, CDW, structural or magnetic transitions \cite{kobayashi_KROPRB2009, FOO_KRO2004}. 
The expected changes due to possible chemical potential shift, to maintain symmetric distribution of the electron-hole excitations around $E_F$, are comparatively smaller than the experimentally observed changes (see appendix~\ref{sec:appendixB}). Our detailed analysis following the Altshuler-Aronov theory of disordered systems \cite{ALTSHULER1979115, DDS_PhysRevLett.98.246401, *Reddy_STIO2019, *SBansal_PhysRevMaterials.7.064007} also fails to describe the temperature dependent spectral lineshape of this quasi-1D system (see appendix \ref{sec:appendixC}). In 1D metallic systems, the presence of interaction within the chain leads to TLL lineshape \cite{Ohtsubo_Bi_PRL2015, *PhysRevB.93.085108,*SCHONHAMMER1993225} of the spectral intensity   given by 
    \begin{equation*} \label{eqnTLL}
	I(\epsilon) \propto T^{\alpha} cosh\biggl(\frac{\epsilon}{2}\biggr) \biggl|\Gamma\biggl(\frac{1+\alpha}{2} + i\frac{\epsilon}{2\pi}\biggl)\biggr|^2 f(\epsilon,T)*G
    \end{equation*}
where, $\epsilon$ = $E/k_{B}T$, $\Gamma$ is the complex gamma function, $f(\epsilon,T)$ is the Fermi-Dirac distribution function and $G$ is Gaussian broadening representing experimental resolution (5 meV). {The TLL state is characterized by non-universal exponent $\alpha$ given by $\alpha$ = ($K_{\rho} + K_{\rho}^{-1} - 2$)/4. 
Here, $K_{\rho}$ is called Luttinger parameter which determines density-density correlation function in the TLL framework \cite{JVoit,Giamarchi_1dBook_2003}. $K_{\rho}$ = 1 for non-interacting case and deviates from 1 when interactions are present \cite{JVoit,Giamarchi_1dBook_2003}. 
Such spectral behaviour of the photoemission intensity with  different exponent $\alpha$ have been observed in variety of 1D/quasi-1D systems \cite{Nicholson_PRL2017,PhysRevLett.82.2540,Kang2021}.}
The systems with TLL behaviour follow a universal scaling relation between the scaled photoemission intensity ($I/T^\alpha$) with temperature-normalized energy ($\epsilon$) such that the curves at different temperature coincide for an appropriate value of $\alpha$ \cite{Ohtsubo_Bi_PRL2015, PhysRevLett.103.136401}. We find $\alpha$ = 0.45 $\pm$ 0.05 to be most appropriate for which the scaling behaviour was observed down to 150 K, as shown in Fig. \ref{fig:Fig3}(b). A closer look at around $\epsilon$ = 0 (see inset) suggests that for $T \leq$ 100 K, the spectra deviate from universal scaling behaviour where the appearance of a Fermi cut-off is evident in the raw spectra (see Fig. \ref{fig:Fig3}(a)), suggesting dimensional crossover below 150 K. 
We also show the TLL lineshape overlapped with the spectral intensity in Fig. \ref{fig:Fig3}(c) for 300 K $\leq$ $T$ $\leq$ 150 K. {It is to note that the high temperature TLL behaviour along with lower Hubbard band is uniquely observed in this system.} A good representation of the photoemission spectra at all temperatures could not be obtained with a single value of $\alpha$ (see Note S3 and S4 of SM \cite{SM}). Also, the spectra at low temperatures exhibiting Fermi cut-off could not be described using any value of $\alpha$ (including $\alpha$ = 0), confirming a higher dimensional electronic behavior with non-Fermi liquid like transport. 
These results are commensurate with the transport properties where a dimensional crossover has been observed at around 150 K \cite{kobayashi_KROPRB2009}.

\begin{figure}[t]
    \centering
    \includegraphics[width=.42\textwidth, trim= 0 0 0 0,clip]{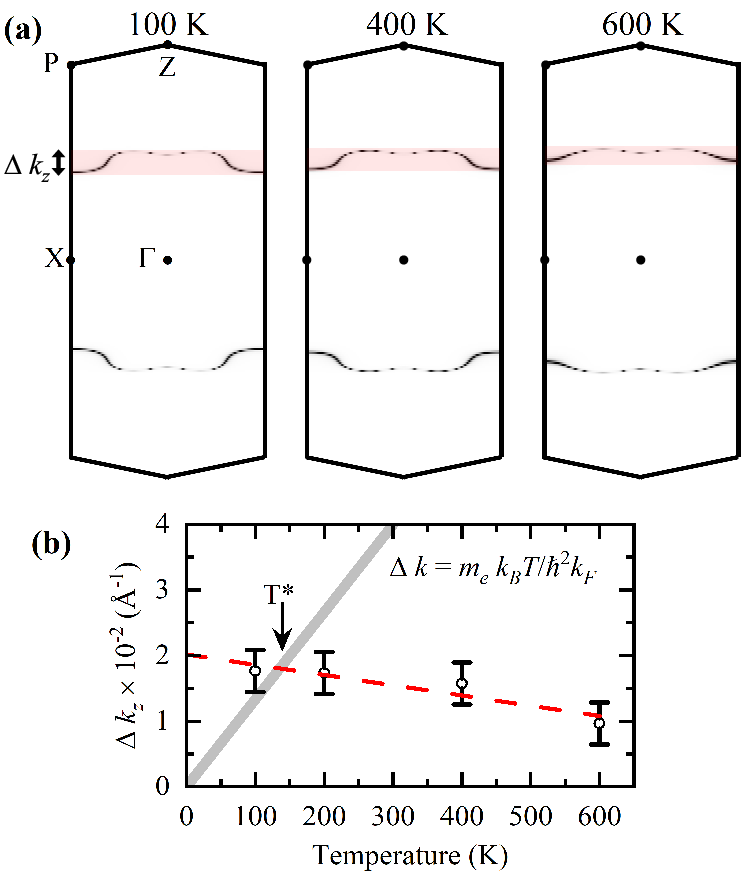}
    \caption{(a) LDA+DMFT calculated FS cut in the $\Gamma$-X-P-Z plane at different temperatures. The FS warping is shown by shaded region. (b) FS warping along $k_z$ as a function of temperature obtained from LDA+DMFT (markers). Red dashed line represents a linear fit. Grey line represents smearing in the $k$-space at Fermi wave vector $k_F$ at temperature $T$. The crossover temperature ($T^*$) is marked by down arrow.} \label{fig:Fig4}
\end{figure} 

Dimensional crossover in quasi-1D systems is associated with interchain electron hoping ($t_{\perp}$) giving rise to warped FS and its competition with thermal energy $k_{B}T$ \cite{Giamarchi_1dBook_2003, silke_PRL2004}.
Below a characteristic temperature $T^* \approx t_{\perp}/k_{B}$, the interchain electron hopping becomes coherent and the 1D character is destroyed, while above $T^*$ the thermal energy dominates to exhibit incoherent transverse hopping leading 1D TLL behaviour \cite{Giamarchi_1dBook_2003, silke_PRL2004}.
Having revealed the dimensional crossover with high-resolution photoemission spectroscopy, we further investigate the effect of temperature on the warping of the FS within the LDA+DMFT framework. 
Figure \ref{fig:Fig4}(a) shows the FS cut in the $\Gamma$-X-P-Z plane, along which the larger FS warping was observed in the full Brillouin zone (Fig. \ref{fig:Fig1}(a)) [see Fig.S4 of SM for FS cut in $\Gamma$-M-Z$_0$-Z plane \cite{SM}].
The Fermi wave vector $k_F$ along $\Gamma$-Z ($k_z$) direction is found to be about 0.08 \AA$^{-1}$ for all the temperatures. As evident, the warping of the FS ($\Delta k_z$) reduces almost linearly with increasing temperature. 
These changes arise due to temperature dependence of the self-energy as well as chemical potential (see appendix \ref{sec:appendixD}). 
The temperature dependent warping with the error bars (the width of the spectral function at $E_F$) is shown with the markers in Fig. \ref{fig:Fig4}(b).
An extrapolate value of $\Delta k_z$ = 0.02 \AA$^{-1}$ at 0 K is obtained \textit{via} a linear fit (red dashed line). In a simplistic approach to compare the thermal energy and the FS warping, we assume a single parabolic band crossing $E_F$, such that $\Delta E/ \Delta k$ = $\hbar^{2}k_{F}/m_e$  with $\Delta E = k_BT$ ($m_e$ is electron mass).
The linear relationship between $\Delta k$ and temperature is shown by grey line in the figure. Schematically,  a crossover temperature ($T^*$) is obtained from this approach and has been marked by a down arrow in Fig. \ref{fig:Fig4}(b). At higher temperatures, thermal energy induced momentum broadening is larger than FS warping, leading to an effectively 1D electronic behaviour, while at temperature below $T^*$ $\sim$ 150 K, the FS warping dominates and transverse hopping becomes coherent, giving rise to a 3D electronic behaviour. {The increased slope of the grey line with inclusion of enhanced electronic mass $m^{*}_{e}$ (due to electron correlation) would slightly reduce $T^{*}$ \cite{Giamarchi_1dBook_2003, silke_PRL2004}}. 
Spectacular observation of similar crossover temperature obtained with this analysis and temperature dependent high-resolution photoemission spectra opens up the gateway to explore the physics of dimensional crossover and related phenomena in various quasi-1D systems using combined theoretical and experimental approaches.
The employed single-site LDA+DMFT approach is insufficient to fully capture the subtleties of one-dimensional systems and more sophisticated theoretical method is required to describe the 1D physics and the crossover from TLL to Fermi or non-Fermi liquid behavior.  Further experiments such as angle-resolved photoemission spectroscopy along with various many body calculations may provide further understanding of the momentum dependence of the Fermi surface, hopping interactions, and effect of non-local correlation in this system.           

\section{Conclusion}
In conclusion, the valence band photoemission spectra in conjunction with LDA+DMFT calculations for K$_{2}$Ru$_{8}$O$_{16}$ reveal moderate strength of electron correlation as manifested by renormalized Ru 4$d$ band along with the signature of lower Hubbard band. Temperature dependent high-resolution photoemission spectra near $E_F$ provides direct evidence of dimensional crossover at $\sim$ 150 K, above which the spectral intensity is well described by TLL lineshape as expected for 1D conductors. The low temperature spectra exhibiting Fermi cut-off with continuously decreasing intensity at $E_F$ suggest a non-Fermi liquid behaviour.  
Fermi surface warping obtained within LDA+DMFT framework increases with lowering temperature and dominate over thermal energy induced momentum broadening at a similar crossover temperature.
The presented findings deepen our understanding of the temperature evolution of the electronic structure across dimensional crossover and motivate further investigations in various quasi-1D systems.

\section*{ACKNOWLEDGMENTS}
We thankfully acknowledge S. Gangadharaiah, IISER Bhopal for fruitful discussions. We acknowledge the support of central instrumentation facility and high performance computing facility at IISER Bhopal. Support from DST-FIST (Project No. SR/FST/PSI-195/2014(C)) is thankfully acknowledged.

\appendix
\section{\label{sec:appendixA} LDA+DMFT spectral function: Varying Hubbard $U$, Hund's $J$ and temperature}
\begin{figure}[h]
        \centering
        \includegraphics[width=.48\textwidth]{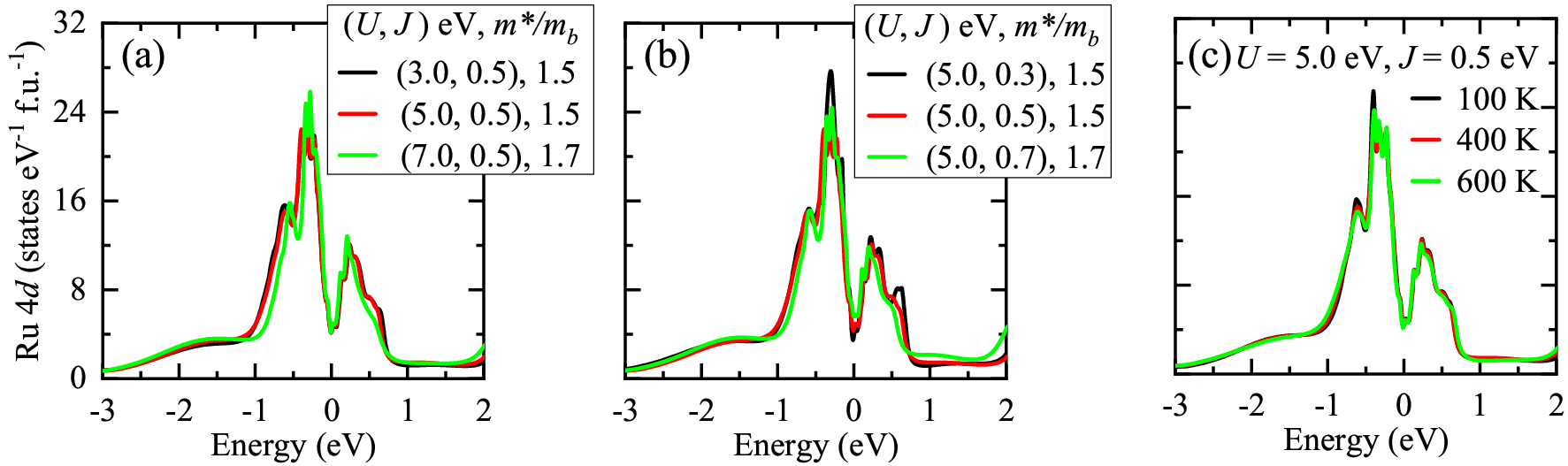}
     \caption{Ru 4$d$ spectral functions obtained from LDA+DMFT calculation for 400 K with (a) varying $U$ (for $J$ = 0.5 eV) and (b) varying $J$ (for $U$ = 5.0 eV). (c) Temperature dependent Ru 4$d$ spectral function for $U$ = 5.0 eV and $J$ = 0.5 eV.} \label{fig:FigS5}
    \end{figure}

The Ru 4$d$ spectral function remains very similar with respect to change in both $U$ (for $J$ = 0.5 eV) and $J$ ($U$ = 5.0 eV) as shown in Fig. \ref{fig:FigS5} (a) and (b), respectively. The calculated mass enhancement factor $m^*/m_b$ does not change drastically and remains 1.6 $\pm$ 0.1. Mass enhancement factor along with strength and position of the incoherent feature also does not change with varying temperature, as shown in Fig. \ref{fig:FigS5}(c).     

\section{\label{sec:appendixB} Spectral changes in the case of asymmetric DOS}

In the case of a symmetric DOS around $E_F$, such as the flat DOS for silver (Ag), the number of thermally excited electrons and holes would be the same, and the chemical potential is expected to remain same with change in temperature. This is evident from the photoemission spectra of Ag, which exhibit no change in the intensity at $E_F$. However, in the case of an asymmetric DOS, the chemical potential might shift with temperature to maintain the filling and symmetric electron-hole excitations. The direction of this shift would depend on the structure of the DOS in the vicinity of $E_F$. 
\begin{figure}[t]
        \centering
        \includegraphics[width=.49\textwidth]{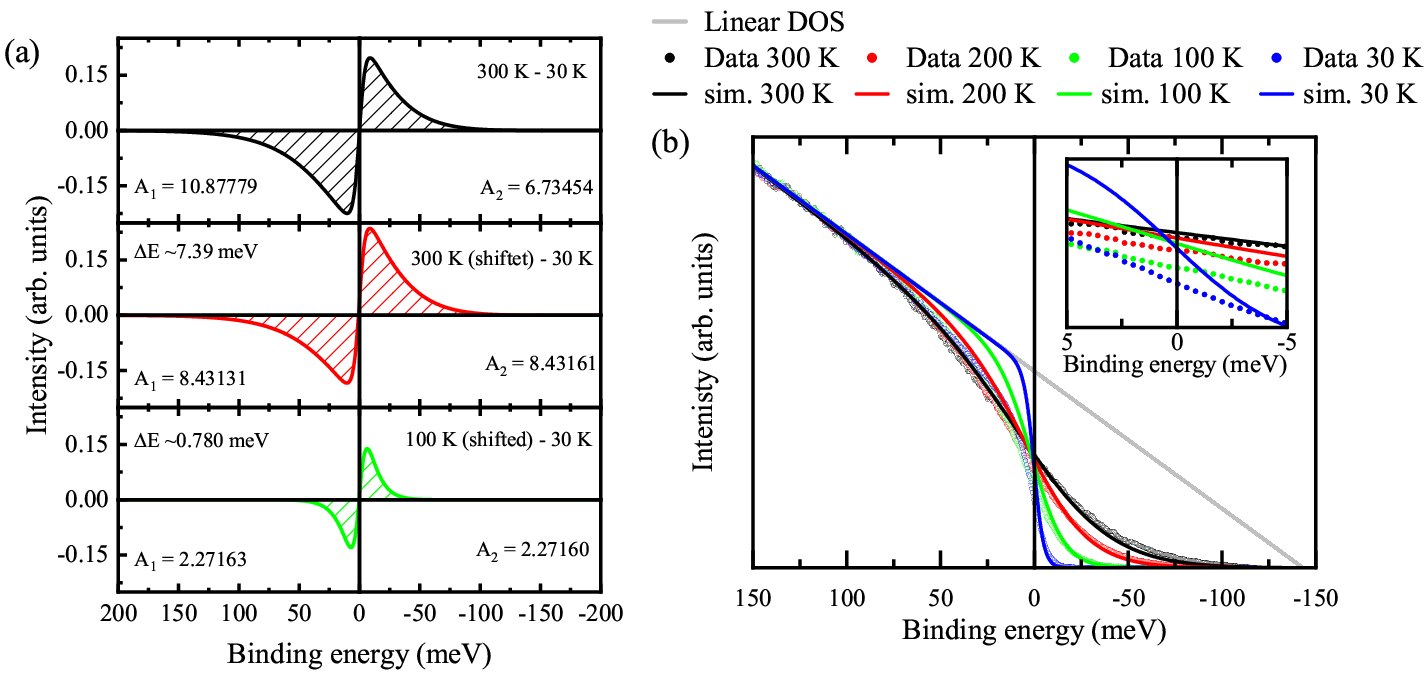}
     \caption{(a) (top panel) Difference of the simulated photoemission spectra at 300 K and 30 K. Difference in simulated spectra after shifting (middle panel) 300 K and (bottom panel) 100 K spectra by $\Delta$E.  A1 (A2) represents the shaded area in occupied (unoccupied) regions. (b) The temperature evolution of the simulated spectra (lines) after accounting for shift in the chemical potential. Markers shows the temperature dependent photoemission spectra of and grey line shows the linear DOS. Inset shows the experimental (markers) and simulated (lines) spectra in the vicinity of $E_F$.} \label{fig:FigR3}
\end{figure} 
To depict the decreasing spectral weight towards $E_F$, we assume a linearly decreasing DOS as shown by the grey line in Fig. \ref{fig:FigR3}(b). The line matches with the spectra at higher binding energy ($>$ 75 meV). The photoemission spectra can be simulated by multiplying the linear DOS with the Fermi-Dirac distribution function at respective temperatures. At a finite temperature, the thermally excited holes in the occupied region would be more than the thermally excited electrons in the unoccupied regions. By subtracting spectra at two temperatures, we can estimate the increase in thermally excited holes and electrons represented by shaded area in the occupied region (A1) and unoccupied region (A2), respectively, as shown in Fig. \ref{fig:FigR3}(a). The asymmetric distribution of electron-hole excitations can be seen where A1 $>$ A2 is obtained using simulated spectra at 300 K and 30 K, as shown in the top panel of Fig. \ref{fig:FigR3}(a). To achieve a symmetric distribution (i.e., A1 = A2), the 300 K spectra is shifted by about 7.39 meV towards the occupied region which represents the shift of chemical potential. A similar analysis is also performed for 100 K (see bottom panel in Fig. \ref{fig:FigR3}(a)).
We next compare these simulated spectra after accounting for the required shift with the experimental spectra in Fig. \ref{fig:FigR3}(b). A reasonable representation of the 300 K experimental spectra (black symbols) is obtained, but a strong deviation is observed at lower temperatures. The decreasing intensity at $E_F$ with lowering temperature is evident for the simulated curves (lines), although, the decrease is significantly smaller than those observed in the experiment spectra (symbols). This can be clearly seen in in the inset of Fig. \ref{fig:FigR3}(b). Overall, the temperature dependence of simulated curves with inclusion of electron-hole symmetricity effect do not agree well with the experimental results (Fig. \ref{fig:Fig3}(a)). Thus the present analysis suggests that only chemical potential shift cannot account for the anomalous behaviour of the photoemission spectra.

\section{\label{sec:appendixC} Photoemission spectra compared with lineshape for disordered systems}

\begin{figure}[t!]
        \centering
        \includegraphics[width=.49\textwidth]{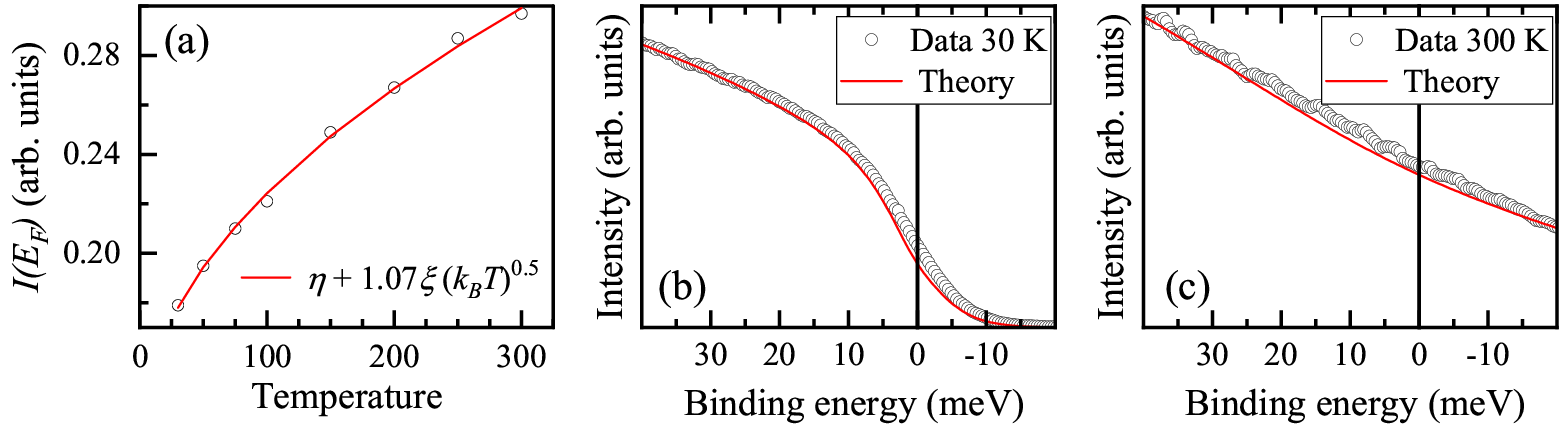}
     \caption{(a) Variation of spectral intensity at $E_F$, $I(E_F)$ with temperature, \textit{T}. The red line shows $\sqrt{T}$ behaviour. Simulated spectral lineshape (red lines) from Althshuler-Aronov theory has been compared with the photoemission spectra (markers) at (b) 30 K and (c) 300 K.} \label{fig:SM_AATheory}
\end{figure}

Monotonous decrease of the spectral intensity at $E_F$ and spectral lineshape in the vicinity of $E_F$ for correlated disordered systems follow Altshuler-Aronov theory \cite{ALTSHULER1979115,DDS_PhysRevLett.98.246401, SBansal_PhysRevMaterials.7.064007, Reddy_STIO2019}. We analyzed our high-resolution photoemission spectra considering the spectral form given by \cite{DDS_PhysRevLett.98.246401}: 	$I(E, T) = [\eta + \xi \{E^2 + (1.07)^4 (k_{B}T)^2\}^{1/4}]f(E,T) $, where $f(E,T)$ is the Fermi-Dirac distribution function, $\eta$ and $\xi$ are fitting parameters. The above equation suggest $\sqrt{T}$ behavior of spectral intensity at $E_F$ and a fitting following the relation $I(E_F) = \eta + 1.07 \xi\sqrt{k_{B}T}$ as shown in Fig. \ref{fig:SM_AATheory}(a) provides a reasonable agreement. It is to note that the anomalous exponent of 0.45 $\pm$ 0.05 is obtained for TLL state at higher temperatures as discussed later as well as in the main text. $\eta$ and $\xi$ were obtained from the fitting and using these parameters, the spectral intensity is simulated and have been compared with the photoemission spectra, as shown in Fig. \ref{fig:SM_AATheory} (b) and (c) for 30 K and 300 K, respectively. The simulated spectra show deviation from the photoemission spectra at both temperatures within $\pm~k_BT$ energy range. Based on this analysis, the observed spectral evolution in K$_{2}$Ru$_{8}$O$_{16}$ can not be attributed to the disorder effect.

\section{\label{sec:appendixD} Temperature dependence of the LDA+DMFT self-energy}
\begin{figure}[h]
        \centering
        \includegraphics[width=.48\textwidth]{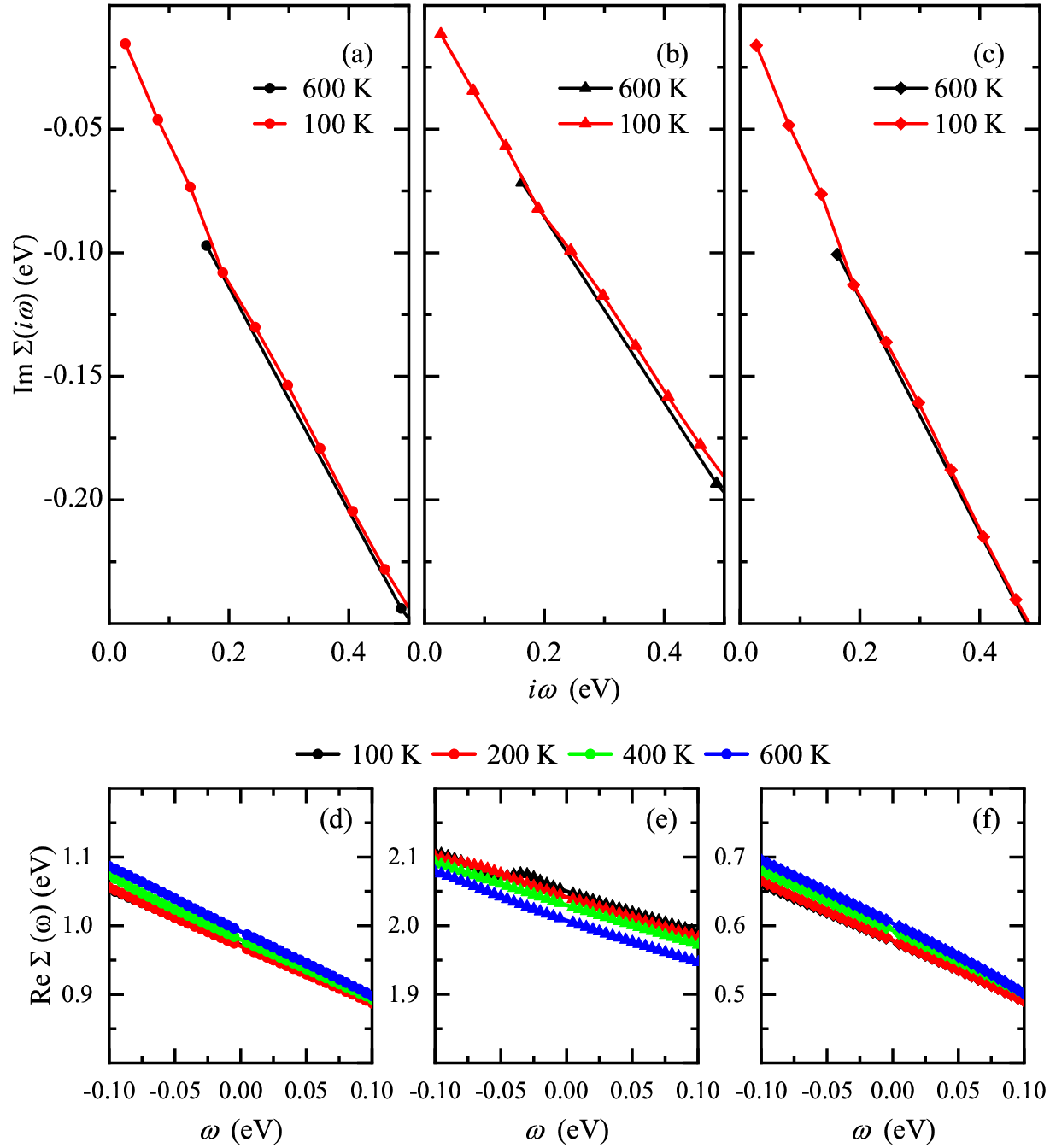}
     \caption{(a-c) Temperature dependence of the imaginary part of the self-energy on Matsubara frequency axis for three Ru~4$d$ ~$t_{2g}$ orbitals. (d-f) Temperature dependence of the real part of the self-energy on real frequency axis for three Ru~4$d$~$t_{2g}$ orbitals. } \label{fig:self-energy}
    \end{figure}

The finite changes in the self-energy and chemical potential modifies the overall spectral function in the vicinity of the Fermi level (see equation \ref{eq:1}), leading to effective temperature dependence of the Fermi surface. The self-energy of the Ru~4$d$~$t_{2g}$ orbitals at two different temperatures on the Matsubara frequency axis exhibit finite changes as shown in Fig. \ref{fig:self-energy} (a-c). These changes can be further observed in the real part of the self energy on the real frequency axis as shown in Fig. \ref{fig:self-energy} (d-e) for different temperatures. We also observed a small shift (less than ~0.01 eV towards higher energy) in the chemical potential while going from 100 K to 600 K in the LDA+DMFT calculations.   

\end{document}


\title{Supplemental Material for ``Origin of dimensional crossover in quasi-one-dimensional hollandite K$_{2}$Ru$_{8}$O$_{16}$"
	}
 
\author{Asif Ali}
\affiliation{Department of Physics, Indian Institute of Science Education and Research Bhopal, Bhopal 462066, India}

\author{Sakshi Bansal}
\affiliation{Department of Physics, Indian Institute of Science Education and Research Bhopal, Bhopal 462066, India}

\author{B. H. Reddy}
\affiliation{Department of Physics, Indian Institute of Science Education and Research Bhopal, Bhopal 462066, India}%

\author{Ravi Shankar Singh}
\email{rssingh@iiserb.ac.in}
\affiliation{Department of Physics, Indian Institute of Science Education and Research Bhopal, Bhopal 462066, India}

\maketitle
    
\subsection*{Note S1: Details of the cRPA calculation}
    We have performed constrained random phase approximation (cRPA) \cite{PhysRevB.80.155134} calculation as implemented in VASP \cite{PhysRevB.47.558, PhysRevB.54.11169,PhysRevB.59.1758} to estimate the Hubbard $U$ and Hund's $J$. A plane wave cut off energy of 550 eV and $k$-grid of 7 $\times$ 7 $\times$ 7 was used. For cRPA, polarization function was evaluated by constructing Ru 4$d$ and O 2$p$ maximally localized wannier functions \cite{PhysRevB.80.155134} using WANNIER90 program \cite{Mostofi2014-za} and $\sim$ 570 empty bands were included. The screened $U$ = 3.33 eV and $J$ = 0.43 eV were obtained for the static limit ($\omega$ = 0) of the cRPA interactions and only Ru 4$d$ orbitals were considered in the correlated subspace.

\subsection*{Note S2: Crystal structure and sample characterization}   
  \begin{figure}[b]
    	\centering
    	\includegraphics[width=.55\textwidth]{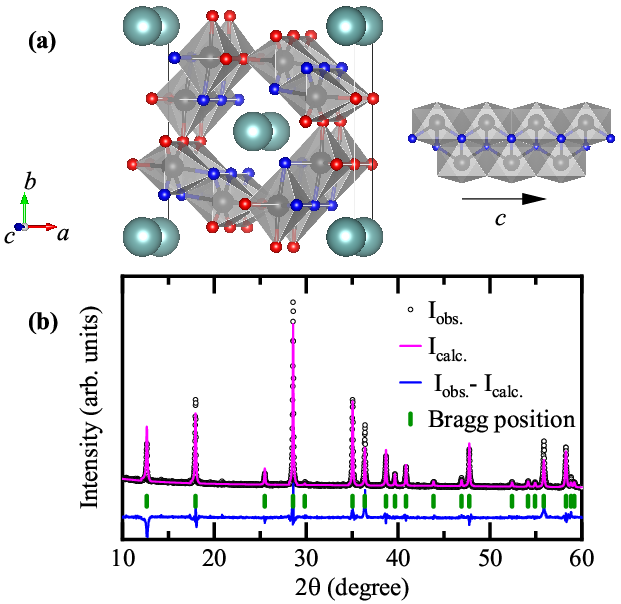}
    	\caption{(a) Crystal structure of K$_{2}$Ru$_{8}$O$_{16}$ depicting tetragonal unit cell. Teal, grey, blue and red spheres represents K, Ru, O1 and O2 atoms, respectively. RuO$_2$ zig-zag chain along $c$-axis is also presented. (b) Room temperature $x$-ray diffraction pattern (black circle) of  K$_2$Ru$_{8}$O$_{16}$ along with the refinement result (pink line). Blue line shows the difference and green bars shows the position of Bragg peaks.} \label{fig:FigS1}
    \end{figure} 

    Figure \ref{fig:FigS1}(a) shows the tetragonal unit cell ($I4/m$; space group No. 87) of K$_{2}$Ru$_{8}$O$_{16}$ and the zig-zag chain along $c$-axis. All the Ru atoms are at crystallographically equivalent positions while oxygens have two different crystallographic positions. Within the zig-zag chain the RuO$_6$ octahedra are connected by edge-shared oxygen O1 (blue spheres), while the interchain connections are via corner shared oxygen O2 (red spheres) of the octahedra.
    The $x$-ray diffraction  pattern and it's Rietveld refinement are shown in Fig. \ref{fig:FigS1}(b) which confirms the phase purity of the sample. Result of the Rietveld refinement reveals tetragonal crystal structure (space group: $I4/m$ (No. 87)) with lattice parameter $a$ = $b$ = 9.8598(1) \AA ~and $c$ = 3.1353(1) \AA, in close agreement with literature \cite{FOO_KRO2004, kobayashi_KROPRB2009}.

\subsection*{Note S3: Tomonaga-Luttinger liquid (TLL) lineshape for various values of anomalous exponent $\alpha$}
\begin{figure}[h]
        \centering
        \includegraphics[width=.77\textwidth]{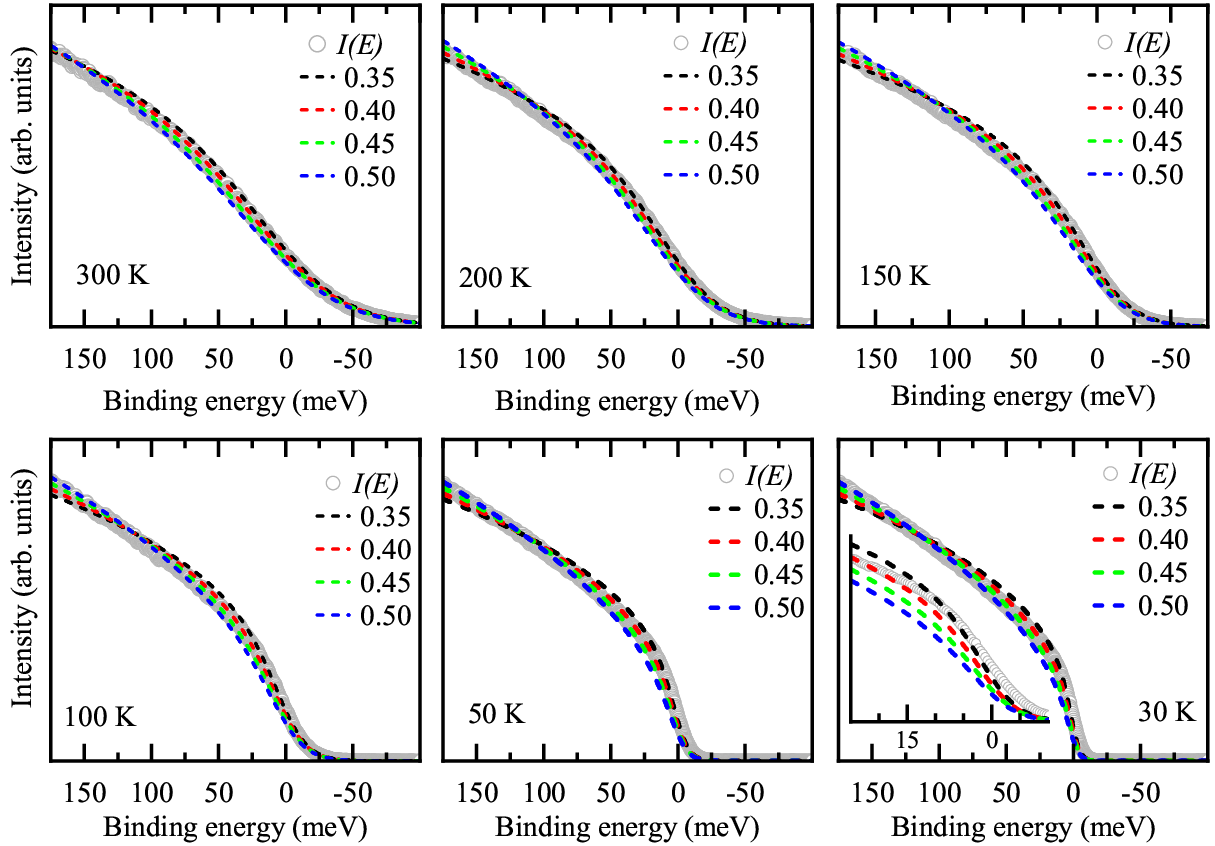}
     \caption{Comparison of photoemission spectra at various temperatures with TLL lineshape simulated for different values of $\alpha$.  30 K spectra shown in small energy window around $E_F$, shown in inset, confirms the non-reproduciblity of the spectra for any value of $\alpha$.} \label{fig:Fig_TLLalpha}
    \end{figure}

The TLL lineshape for various value of $\alpha$ is compared with photoemission spectra at different temperature in Fig. \ref{fig:Fig_TLLalpha}. The simulated curve with $\alpha$ = 0.45 $\pm$ 0.05 shows good agreement for the spectra at higher temperatures while at temperature below 150 K as shown in top panel. The low temperature spectra could not be well fitted for any value of $\alpha$ as shown in bottom  panel.  

\subsection*{ Note S4: Universal scaling behaviour of the photoemission spectra }
\begin{figure}[h]
        \centering
        \includegraphics[width=.8\textwidth]{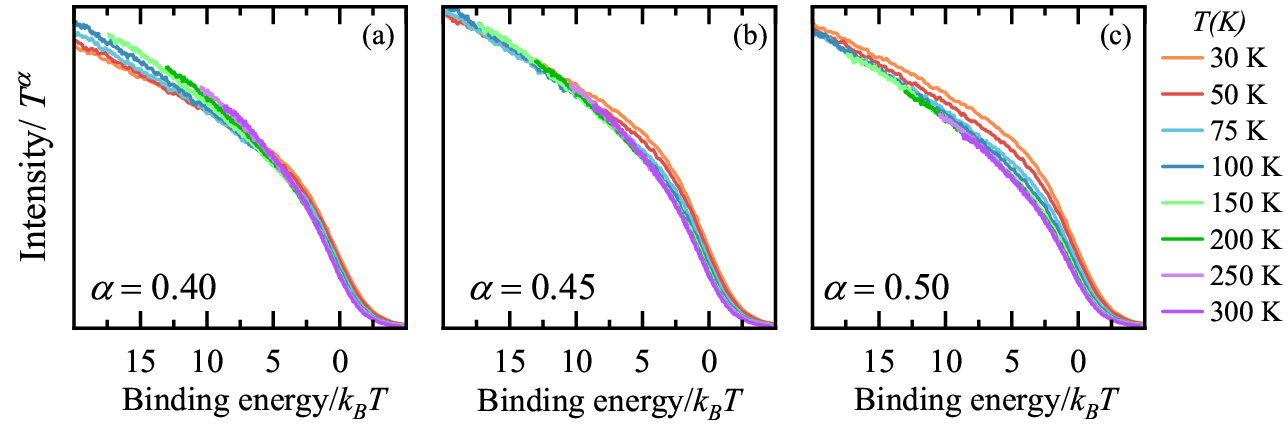}
     \caption{Universal scaling plot of photoemission spectra at different temperatures for different values of $\alpha$} \label{fig:scaledalpha}
    \end{figure}

The universal scaling plots of the photoemission spectra for different $\alpha$ = 0.40, 0.45 and 0.50 are shown in Fig. \ref{fig:scaledalpha} (a), (b) and (c), respectively. Figure \ref{fig:scaledalpha}(b) suggests a good scaling behaviour for temperatures above 150 K for $\alpha$ 
= 0.45 and has also been shown in the main text.  

\subsection*{ Note S5: FS cut in the $\Gamma$-M-Z$_0$-Z plane obtained from LDA+DMFT}
\begin{figure}[h]
        \centering
        \includegraphics[width=.5\textwidth]{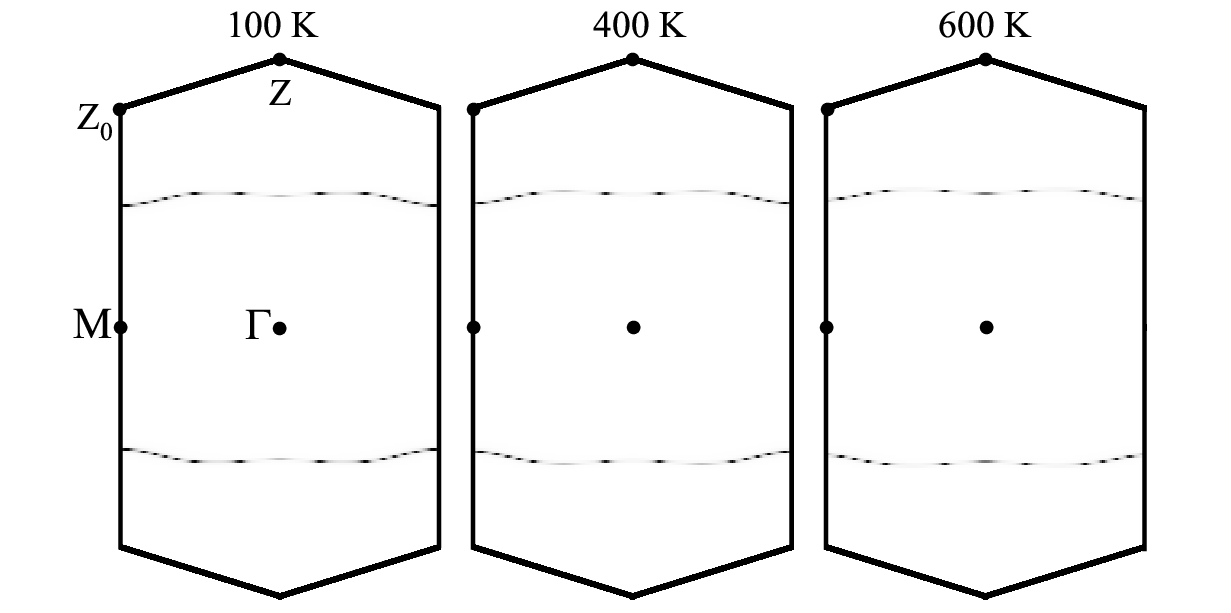}
     \caption{FS cut obtained in $\Gamma$-M-Z$_0$-Z plane at different temperature from the LDA+DMFT calculations.} \label{fig:FigS7}
    \end{figure}

Weak warping of the LDA+DMFT FS in $\Gamma$-M-Z$_0$-Z plane, shown in Fig. \ref{fig:FigS7}, exhibits temperature independent behaviour. 
